\documentclass[useAMS,usenatbib]{mn2e}

\usepackage{natbib, aas_macros}
\usepackage{lmodern}
\usepackage{cite}
\usepackage[dvips]{graphicx}
\usepackage{wrapfig}
\graphicspath{{image/}}
\usepackage{color}
\usepackage{amsmath}
\usepackage{amssymb}
\usepackage{rotating}
\usepackage{appendix}

\bibpunct{(}{)}{;}{a}{}{,} 

\newcommand{\HI}{\rm H~{\sc i}}

\newcommand{\HeI}{\rm He~{\sc i}}
\newcommand{\HeII}{\rm He~{\sc ii}}
\newcommand{\TB}{\delta T_{\rm b}}
\newcommand{\MSUN}{\rm M_{\odot}}
\newcommand{\XHI}{x_{\rm HI}}
\newcommand{\XHII}{x_{\rm HII}}

\newcommand{\TS}{T_{\rm S}}
\newcommand{\TK}{T_{\rm K}}
\newcommand{\TCMB}{T_{\gamma}}
\newcommand{\lya}{\rm {Ly{\alpha}}}

\newcommand{\OmegaB}{\Omega_{\rm B}}
\newcommand{\Omegam}{\Omega_{\rm m}}

\def\DEL2{\Delta^2}

\title[Light-cone effect on cosmic dawn 21-cm signal]{21-cm signal from cosmic dawn - II: Imprints of the light-cone effects}

\author[Ghara, Datta \& Choudhury]{Raghunath Ghara$^1$\thanks{Email: raghunath@ncra.tifr.res.in}, Kanan K. Datta$^{2}$\thanks{Email: kanan.physics@presiuniv.ac.in} and T. Roy Choudhury$^1$\thanks{Email: tirth@ncra.tifr.res.in} \\
$^1$ National Centre for Radio Astrophysics, TIFR, Post Bag 3, Ganeshkhind, Pune 411007, India\\ 
$^2$ Department of Physics, Presidency University, 86/1 College Street, Kolkata - 700073, India } 

\begin{document}

\date{Accepted ?; Received ??; in original form ???}

\pagerange{\pageref{firstpage}--\pageref{lastpage}} \pubyear{?}

\maketitle

\label{firstpage}


\begin{abstract}
Details of various unknown physical processes during the cosmic dawn and the epoch of reionization can be extracted from observations of the redshifted 21-cm signal. These observations, however, will be affected by the evolution of the signal along the line-of-sight which is known as the ``light-cone effect''. We model this effect by post-processing a dark matter $N-$body simulation with an 1-D radiative transfer code. We find that the effect is much stronger and dramatic in presence of inhomogeneous heating and $\lya$ coupling compared to the case where these processes are not accounted for. One finds increase (decrease) in the spherically averaged power spectrum up to a factor of 3 (0.6) at large scales ($k \sim 0.05\, \rm Mpc^{-1}$)  when the light-cone effect is included, though these numbers are highly dependent on the source model. The effect is particularly significant near the peak and dip-like features seen in the power spectrum. The peaks and dips are suppressed and thus the power spectrum can be smoothed out to a large extent if the width of the frequency band used in the experiment is large. We argue that it is important to account for the light-cone effect for any 21-cm signal prediction during cosmic dawn.

\end{abstract}


\section{Introduction} 
\label{intro}

Understanding the physical processes during the `cosmic dawn',  when the very first sources formed in the universe, and the epoch of reionization (EoR), when the primordial neutral hydrogen (\HI) got ionized, are some of the major goals of present day observational astronomy. The existing probes like observations of cosmic microwave background (CMB) \citep{Komatsu11, Planck2013, 2015arXiv150201589P} and high redshift quasar absorption spectra \citep{Gunn65, becker01, fan03, fan2006, Goto11} suggest that the process of reionization occurred mainly during the redshift range $6 < z < 15$ \citep{ Malhotra06, Choudhury06a, mitra2011, mitra2012, 2015arXiv150202024R}. However, these observations are still unable to constrain the details of the reionization very accurately, e.g., the exact duration of reionization, nature of the reionization process, nature of sources present during those epochs, ionization and thermal states of the intergalactic medium (IGM) etc.

Observations of redshifted \HI ~21-cm signal from these epochs promise to provide us with the details of many such unknowns \citep[see][for details]{Furlanetto2006, Morales10,Pritchard12}. This signal not only carries information about the ionization state of the hydrogen in the IGM, but also the thermal state of the IGM. The first generation low frequency radio telescopes, like Low Frequency Array (LOFAR)\footnote{http://www.lofar.org/} \citep{van13}, the Precision Array for Probing the Epoch of Reionization (PAPER)\footnote{http://eor.berkeley.edu/} \citep{parsons13}, the Murchison Widefield Array (MWA)\footnote{http://www.mwatelescope.org/} \citep{bowman13, tingay13}, the Giant Metrewave Radio Telescope (GMRT) \footnote{http://www.gmrt.tifr.res.in}\citep{ghosh12, paciga13},  have already started observations with the aim to detect the 21-cm signal. These instruments are focussing mainly on the EoR as they do not contain the lower frequency bands required to probe the cosmic dawn. The next generation telescopes like the Square Kilometre Array (SKA)\footnote{http://www.skatelescope.org/}, which are equipped with lower frequency detectors and extremely sensitive, should be able to probe the cosmic dawn and reveal the unknown properties of the very first sources in the Universe \citep{mellema13}. 

The 21-cm signal from \HI ~is sensitive to various quantities such as the number density and clustering of sources, their ionizing, heating and coupling efficiencies, the escape fraction of photons at various frequency bands, various feedback effects etc. These quantities are eventually parametrized in terms of the collapse fraction of dark matter haloes,  fraction of baryons converted into stars, the stellar initial mass function, $\lya$, UV and X-ray luminosities, X-ray spectral index etc  \citep{Morales10, Pritchard12, Mesinger2014}. In addition, line of sight effects such as the peculiar velocities and evolution of the 21-cm signal could also affect the signal significantly. It is thus very important to model the expected \HI ~21-cm signal properly in order to interpret the observations. A large amount of theoretical modelling and simulations is needed to explore all possible scenarios and such analyses are expected to play a major part in  designing 21-cm experiments. Different approaches such as analytical \citep[e.g.,][]{furlanetto04, 2014MNRAS.442.1470P}, semi-numerical \citep{zahn2007, mesinger07, santos08, Thom09, choudhury09, plante13, ghara14}, and numerical \citep{Iliev2006,  mellema06, McQuinn2007, shin2008, baek09}  have been attempted in modelling the signal.

In this paper we investigate one of the important line of sight effects i.e, the `light-cone effect' on the 21-cm signal from the cosmic dawn and the EoR. While simulating the signal, it is generally assumed that every part of a simulation box has the same redshift. We call each such simulation snapshot as `coeval box'.  In reality, regions which are nearer to the observer along a line of sight will have lower redshifts than the regions which are far away. As a result the observed signal will have effects of redshift evolution imprinted on it. Studies through analytical modelling and simulations have been done to understand this so called `light-cone' effect on the two point correlation function \citep{barkana2006, zawada14}  and power spectrum \citep{Datta2012b, plante13, datta2014b}. \citet{datta2014b} find that, depending on the observational bandwidth, the effect could enhance the power spectrum by a factor of up to $\sim 5$ at the initial stages of the EoR and suppress by a significant amount at the last stages of the EoR. Interestingly, no significant light-cone anisotropy has been found in \citet{datta2014b}.  It has also been noticed that the light-cone effect on the power spectrum will be significant when it evolves non-linearly with redshift as the linear evolution is smoothed out \citep{Datta2012b}. However, these studies assume that the entire IGM is always heated significantly higher than the CMB brightness temperature and that the $\lya$ coupling is very strong. These assumptions may hold during the later stages of reionization but may not be true at the cosmic dawn and the initial phase of reionization. It is believed that the inhomogeneities in heating and $\lya$ coupling will significantly influence the signal during cosmic dawn and early phase of the EoR (when the mass averaged ionization fraction $\XHII \lesssim 0.2$). These make the evolution of the \HI ~power spectrum with redshift more dramatic and thus we expect a very strong light-cone effect on the signal. \citet{zawada14} too have studied this using numerical simulations with the main focus on the two point correlation functions. In this work, we self consistently calculate the $\lya$ coupling, heating of the IGM for various source models and reionization histories and study the light-cone effect on the \HI ~power spectrum.

We have organized the paper in the following way. In section \ref{simu}, we have briefly described the procedure to generate the brightness temperature maps of the 21-cm signal self-consistently using an 1-D radiative transfer code and the method to incorporate the light-cone effect in the coeval box. We have presented our results in section \ref{res} before summarizing and discussing our results in section \ref{conc}.  The cosmological parameters used for the simulation are $\Omegam=0.32$, $\Omega_\Lambda=0.68$, $\OmegaB=0.049$, $h=0.67$, $n_{\rm s}=0.96$, and $\sigma_8=0.83$,  consistent with the recent results of $Planck$ mission \citep{Planck2013}.

\section{Simulations}
\label{simu}
\subsection{The 21-cm signal}
\label{21_cm}

The redshifted 21-cm signal is expected to be measured as an offset from the CMB radiation. The differential brightness temperature, observed at a frequency $\nu_{\rm obs}$ along a line of sight $\mathbf{\hat{n}}$, can be written as \citep[e.g,][]{Furlanetto2006}

\begin{eqnarray}
\TB (\nu_{\rm obs}, \mathbf{\hat{n}}) \!\!\!\! &\equiv& \!\!\!\! \TB (\mathbf{x}) = 27 ~ \XHI (z,\mathbf{x}) [1+\delta_{\rm B}(z,\mathbf{x})] \left(\frac{\OmegaB h^2}{0.023}\right) \nonumber\\
&\times& \!\!\!\!\left(\frac{0.15}{\Omegam h^2}\frac{1+z}{10}\right)^{1/2}\left[1-\frac{\TCMB(z)}{T_{\rm S}(z,\mathbf{x})}\right]\,\rm{mK},
\nonumber \\
\label{brightnessT}
\end{eqnarray}
where $\bf{x}$ is the distance along the line of sight to redshift $z$, which can be denoted as $1+z = 1420~{\rm MHz}/\nu_{\rm obs}$. The quantities $\XHI(z,\mathbf{x})$ and $\delta_{\rm B}(z,\mathbf{x})$ denote the neutral hydrogen fraction and the density contrast in baryons respectively at point $\mathbf{x}$ and redshift $z$. The quantity $\TS$ is the spin temperature of the neutral hydrogen gas in the IGM, which is determined by the coupling of the neutral hydrogen gas with the CMB photons by Thomson scattering, $\lya$ coupling \citep{wouth52} and collisional coupling \citep{field58,Furlanetto06a}. The quantity $\TCMB(z)$ = 2.73 $\times (1+z)$ K is the CMB brightness temperature at redshift $z$. Note that we have not considered any effect from the peculiar velocities  \citep[e.g,][]{bharadwaj04, barkana05a} of the gas in the IGM in the above equation.

The first generation radio telescopes are not sufficiently sensitive to image the differential brightness temperature of the 21-cm radiation. Instead these are expected to measure the signal in terms of statistical quantities like the power spectrum of $\TB$ fluctuations.  We have presented our results in terms of the dimensionless power spectrum $\Delta^2(k)=k^3P(k)/2\pi^2$ which also represents the power per unit logarithmic interval in scale $k$. Here $P(k)$ is the spherically averaged power spectrum of $\TB$ fluctuations, which is defined as
\begin{equation}
\langle \hat{\TB}(\mathrm{\mathbf{k}}) \hat{\TB}^{\star}(\mathbf{k'})\rangle = (2 \mathrm{\pi})^3 \delta_D(\mathbf{k - k'}) P(k),
\label{ps}
\end{equation}
where $\hat{\TB}(\mathrm{\mathbf{k}})$ is the Fourier transform of $\TB(\mathbf{x})$ defined in equation (\ref{brightnessT}). In general the universe is expected to be isotropic at large scales, though several effects can generate anisotropy in the observed signal. Line of sight peculiar velocity of gas (or the so called redshift-space distortion (RSD)) in the IGM can affect the $\TB$ fluctuations and hence the signal along the line of sight and thus can introduce a difference in the power spectrum along and perpendicular to the observed direction \citep{mao12,Majumdar13,Jensen13}. On the other hand the Alcock-Paczynski effect can also significantly contribute to the anisotropy in the observed signal \citep{Alcock1979,ali2005MNRAS.363..251A,2006MNRAS.372..259B}. In general, the anisotropic 21-cm power spectrum is denoted as $\DEL2 (k,\mu)$, where $\mu=\rm \cos\theta$ with $\theta$ being the angle between the line of sight and the Fourier mode ${\bf k}$. Since the light-cone effect too influences the signal only along the line of sight, it can in principle make the signal anisotropic. In this study we mainly focus on the light-cone effect on the spherically averaged power spectrum and we will discuss the anisotropy introduced by this to the 21-cm power spectrum in section \ref{ani_lc}.

The simulation of the $\TB$ signal involves three main steps: (i) generation of underlying baryonic density and velocity fields, (ii) modelling the radiation sources and (iii) computing the propagation of ionization and heating fronts using an 1-D radiation transfer code. The method used in this paper is essentially described in \citet[][hereafter Paper I]{ghara14}. We summarize the main steps in the next few subsections.


\subsection{$N-$body simulations}
\label{nbody_sim}

The density and velocity fields, used in our simulation, are generated using the publicly available code {\sc cubep}$^3${\sc m}\footnote{\tt http://wiki.cita.utoronto.ca/mediawiki/index.php/CubePM} \citep{Harnois12} which is a massively parallel hybrid  particle-particle-particle-mesh (P$^3$M) code. The simulation started at redshift $z=200$ with the initialization of the particle positions and velocities using {\sc camb} transfer function\footnote{\tt http://camb.info/}  \citep{lewis00} and employing Zel'dovich approximation. 

The properties of the simulations we have used in this work are: (i) the number of particles used is $1728^3$, (ii)  the box size is 200 $h^{-1}$ comoving Mpc (cMpc), (iii) the number of grid points used are  $3456^3$ and (iv) the mass of each dark matter particle is $2 \times 10^8\,  \MSUN$. Since the position and velocity arrays of  the simulation are too large in size, we prefer to generate the density and  velocity fields on a grid 8 times coarser than the simulation grid using a top-hat filter. These snapshot files are generated between $25 \geq z \geq 6$ in equal time gap of $10^7$ years. The baryons are assumed to be simply tracing the dark matter, i.e., if dark matter density at position  $\mathbf{x}$ is $\rho_{\rm DM}(\mathbf{x})$, then the baryonic density will be $\rho_{\rm B}(\mathbf{x}) = \left(\OmegaB/\Omegam \right) \times \rho_{\rm DM}(\mathbf{x})$.  The velocity files on the coarse grid are used to incorporate the peculiar velocity effects in the 21-cm brightness temperature.  

For each snapshot, haloes were identified using a runtime halo finder algorithm which uses spherical over-density method. The minimum halo mass resolved using this method is $\sim 2 \times 10^9\, \MSUN$.  Apart from these haloes, it is expected there will also be considerable number of small mass haloes $\sim 10^8\, \MSUN$ where the gas can cool via atomic transitions and form stars. As the number of such small mass haloes is quite high, these haloes may have significant impact on the reionization scenarios, particularly at the early stages. Identifying such small mass haloes with the spherical over-density method requires simulation box of a higher resolution, which is somewhat beyond our ability right now. Hence we employ a sub-grid method to find the haloes down to masses as small as $\sim 10^8\, \MSUN$. We have followed the extended Press-Schechter model of \citet{Bond1991} and hybrid prescription of \citet{barkana2004} for implementing the sub-grid model (for details please see Paper I).


\subsection{Modelling the sources}
\label{sim_source}

The main sources of ionizing photons are the stars residing in galaxies, which form within dark matter haloes. We assume that haloes having mass $\gtrsim 10^8\, \MSUN$ contribute to ionizing radiation in regions that are neutral. Within ionized regions, the corresponding threshold mass is taken to be $\sim 10^9\, \MSUN$ because of radiative feedback. The stellar mass of a galaxy in a hosting dark matter halo of mass $M_{\rm halo}$ is
\begin{equation}
M_\star=f_\star \left(\frac{\OmegaB}{\Omegam}\right) M_{\rm halo},
\end{equation}
where $f_\star$ is the fraction of baryons residing within the stars in a galaxy. This fraction depends on the metallicity and mass of the galaxy and  is not well constrained for early galaxies. Hence for simplicity we have assumed a constant $f_\star$ throughout the reionization history, with its value is chosen in such a way  that the resulting reionization history is consistent with the constraints obtained from the optical depth measurement of the CMB observations.

The spectral energy distribution (SED) of a galaxy can be calculated using stellar population synthesis codes provided one knows the values of the initial mass function (IMF) of the stars, initial metallicity etc. We assume a Salpeter IMF for the stars with mass range 1 to 100 $\MSUN$ throughout the reionization history. The metallicity evolution is taken from the models of \citet{dayal09a} and \citet{dayal10} which are consistent with the best-fit mass-metallicity relation.
As in Paper I, we generate the UV and near-infrared (NIR) spectral energy distributions of the galaxies for standard star formation scenarios using the code {\sc pegase2}\footnote{\tt http://www2.iap.fr/pegase/} \citep{Fioc97}. The fraction of UV ionizing photons that escape into the IGM from the galaxies ($f_{\rm esc}$) is highly uncertain for high redshift galaxies. For every redshift of interest, we choose the value of $f_{\rm esc}$ so as to obtain a given reionization history, which is explained later in section \ref{res}.

 In this simulation, we have considered ionization of both hydrogen and helium in the IGM by central sources. The photons from the stars can only ionize \HI~ and \HeI, however ionization of \HeII ~and heating of the IGM is not efficient by the stellar sources because of the lack of high energy ($\gtrsim$ 50 eV) photons. Following our previous work (Paper I), we model the X-rays from galaxies as having a power-law SED with a spectral index $\alpha$:
\begin{equation}
I_q(E) = A ~ E^{-\alpha}, 
\end{equation}
where the normalisation constant $A$ can be fixed in terms of the ratio $f_X$ of X-ray to UV luminosity from the source. This kind of a X-ray spectrum is expected from a miniquasar-like source, i.e., a galaxy with intermediate mass black hole of mass in the range of $10^3-10^6\, \rm \MSUN$ \citep{elvis94, laor97, vanden01, vignali03}. There could be other sources of high energy photons like the high mass X-ray binaries \citep{frag1, frag2,Fialkov14} and the hot interstellar medium within early galaxies \citep{Pacucci2014}. However, we concentrate on a power-law X-ray spectrum in this work.

Note that the parameter $f_X$ can be related to the central black hole to galaxy mass ratio. Since main purpose of this paper is to study the light-cone effect for models which incorporate heating and $\lya$ coupling self-consistently, we have kept the values $f_X$ and $\alpha$ to be fixed. We choose $f_X$ to be 0.05 and $\alpha$ to be 1.5 for our study, which are consistent with the observations of local quasars. The effect of choosing other values of $f_X$ and $\alpha$ was studied in detail in Paper I.


\subsection{$\TB$ maps using 1-D radiative transfer}
\label{sim_rt}

As soon as the first sources of light appear in the dark matter haloes, they start to influence the ionization and the thermal state of the IGM in the universe. Since the UV photons have relatively smaller mean free path, they will be absorbed by the neutral hydrogen in the immediate neighbourhood, resulting in highly ionized bubbles around the sources. However, the X-rays can propagate long distances in the IGM and partially ionize and heat up the medium. Although it is necessary to carry out 3-D radiative transfer simulations to generate the ionization and heating maps, it requires huge computational power and long run time. We rather prefer an alternative method based on 1-D radiative transfer which is faster and hence more efficient for exploring the unknown parameter space \citep{Thom09,thomas11}. The method is briefly described in the following, we refer the readers to Paper I for the details.

 (i) The ionization and heating profile around a galaxy depends on the luminosity and the surrounding neutral hydrogen distribution. We generate the ionization and heating profiles around sources for different galaxy masses, redshifts and background gas densities. Though the baryonic distribution around a galaxy changes with distance, but for simplicity we assume the gas is distributed uniformly around the source. This is one of the limitations of our method.

 (ii) Given the list of haloes at a certain redshift, we calculate the corresponding galaxy masses within the haloes as stated in section \ref{sim_source}. We assume that no star formation occurs within low mass haloes ($< 10^8\, \MSUN$). In order to incorporate radiative feedback, we suppress the galaxy formation within newly formed haloes having mass $M < 10^{9}\, \MSUN $ if they form in  already ionized regions, i.e., regions of the IGM with $\XHII$ larger than 0.5. \footnote{ The radiative feedback is effective when the IGM is photo-heated to a temperature $> 10^{4}\, \rm K $, which raises the cosmological Jeans mass and thus suppresses galaxy formation in low mass haloes with mass $M < 10^{9}\, \MSUN $. In general, regions with temperature $> 10^{4}\, \rm K $  are highly ionized (with $\XHII >$  0.5), hence our implementation of feedback should be reasonable. In more realistic scenarios, the feedback is not expected to be so drastic, e.g., $M < 10^9 \MSUN$ haloes within ionized regions can retain a fraction of their gas and possibly form stars at a lower rate. Our method does not account for such effects.}

 (iii) Depending on the mass, background gas distribution and redshift of the galaxies, we estimate the ionizing photons from the galaxies. 
The ionized bubbles around individual sources are computed by solving the radiative transfer equations in 1-D. In case of overlap between the bubbles, we estimate the unused ionizing photons and distribute them equally among the overlapping ionized bubbles.

 (iv) The X-rays can penetrate into  and partially ionize the neutral IGM. Beyond the highly ionized regions around the sources, the ionization fraction is estimated using the pre-generated profiles.  Once ionization maps are generated, the kinetic temperature maps are generated using a correlation of kinetic temperature and ionization fraction in the partially ionized regions (see Paper I for details). 

 (v) As far as the radiative transfer of the $\lya$ photons are concerned, we simply assume the escape fraction of the $\lya$ photons to be 1 and the number density of $\lya$ photons from the galaxies decrease as $1/d^2$, where $d$ be the distance from the source. The $\lya$ coupling coefficient is calculated from the $\lya$ flux, which completes every quantity required for computing the $\TS$ maps. The brightness temperature maps are generated in the simulation box using equation (\ref{brightnessT}).

(vi) The effect of the peculiar velocity of the gas in the IGM is incorporated into the $\TB$ maps using the cell movement method (or Mesh-to-Mesh (MM)-RRM scheme) described in \citet{mao12}, which is time efficient and computationally easy to implement. The details can be found in Paper I.


\subsection{Implementing the light-cone effect}
\label{l_cone}
The light-cone effect essentially accounts for the evolution of the signal with redshift within the simulation box or the so called ``coeval cube'' (CC). \citet{Datta2012b} have described the procedure to incorporate the effect and produce the light-cone cube. In our study we have incorporated the effect in individual coeval cubes instead of generating the full redshift range light-cone cube as done in \citet{Datta2012b}.  We assume that the central ``slice" of the coeval cube represents the cosmological redshift, where a ``slice" is a 2-D map having width $\sim$ 0.7 cMpc (corresponding to one pixel of the simulation box).  Along a given line of sight, different distances represent different redshifts, and thus $\TB$ will be different from that in the coeval cube. The steps to include the light-cone effect are as follows:

(i) We generate the so called ``coeval cubes'' of 21-cm $\TB$ maps at different redshifts following the method stated in section \ref{sim_rt}. Let us assume that we have $N$ numbers of such cubes which correspond to the cosmological redshifts $z_i$, where $i=1,N$, with $z_1 (z_N)$ being the lowest (highest) redshift in our simulation (in our case, we have 77 such coeval cubes between redshift 6 and 20).  

(ii) Each of these coeval cubes have a length $L$ and contain $n^3$ grid points, which in our case have the values 200 $h^{-1}$ cMpc and $432^3$ respectively. We assume the central 2-D slice of the cube, in the perpendicular plane of the line of sight ($x$-axis in this case), correspond to the cosmological redshift of the coeval cube.  The comoving distance $D(z_l,z_u)$ between two redshifts $z_l$ and $z_u$ is given by,
\begin{equation}
D(z_l,z_u)=\int\limits_{z_l}^{z_u} \frac{c}{H(z)} dz,
\label{distz1z2}
\end{equation}
where $c$ is the speed of light and $H(z)$ is the Hubble constant at redshift $z$. Depending on the distance from the central slice along the line of sight, we calculate the corresponding redshift for each slice.  For example, the redshift $z_p$ corresponding to the  $p$th slice is calculated by demanding that $D(z_c,z_p)$ in equation (\ref{distz1z2}) is equal to the distance $(p-n/2)\times L/n$ of the $p$th slice from the central slice. Note that $z_c$ is the redshift corresponding to the central slice.  In the light-cone cube, the signal at each slice would be at the redshift corresponding to that slice and not the cosmological redshift of the coeval cube.

(iii) If the $p$th slice corresponds to a redshift $z_p$, which satisfies the condition $z_i< z_p< z_{i+1}$, then the $\TB$ maps for that slice is computed by linearly interpolating the maps of the slices at redshifts $z_i$ and $z_{i+1}$. Applying the same procedure to all the slices, the ``light-cone cube'' is generated which corresponds to the redshift of the central slice.

Note that the redshift-space distortion needs to be applied to the brightness temperature maps before implementing the light-cone effect, otherwise the redshift evolution of the gas velocities will not be incorporated properly.

\section{Results }
\label{res}

\begin{figure}
\begin{center}
\includegraphics[scale=0.7]{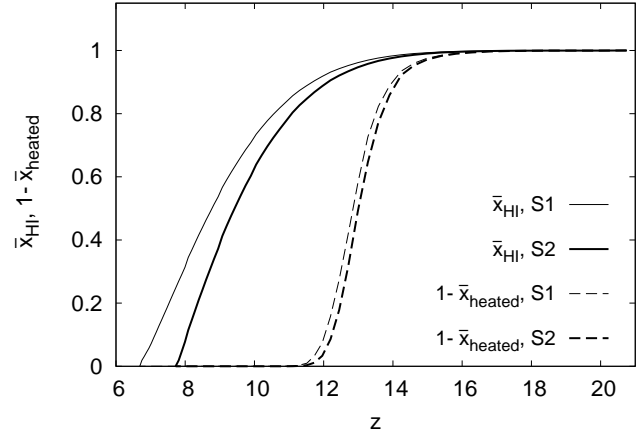}
    \caption{ The solid curves represent the evolution of the mass averaged neutral fraction of hydrogen ($\bar{x}_{\rm HI}$) in the IGM. The dashed curves represent the evolution of ($1-\bar{x}_{\rm heated}$), where  $\bar{x}_{\rm heated}$ is the volume averaged heated fraction of the IGM. The thin and thick curves represent models $S1$ and $S2$ respectively. We have defined `heated region' as a region with $\TK$ larger than $\TCMB$. The X-ray spectrum from the central mini-quasar follows a power law with index $\alpha=$1.5 and the X-ray to UV luminosity of the source is 0.05 for this model.}
   \label{ionfrc}
\end{center}
\end{figure}

In order to compute the $\TB$ maps, one first has to choose appropriate reionization and heating histories, which in turn depend on the nature of sources. In this work, we consider three models of radiation sources $S1, S2$ and $S3$ as listed in Table \ref{tab2} to investigate the effect of different reionization histories on the $\TB$ maps. For all the three source models, we assume each galaxy to contain a mini-quasar like source at the centre which radiates X-rays with a power law distribution having index $\alpha=$1.5, and the X-ray to UV luminosity fraction of the source $f_X$ is 0.05. We have fixed the parameter $f_{\star}$ to 0.1 throughout the reionization history, while the escape fraction $f_{\rm esc}$ is taken to be different in different models. In model $S1$, we take $f_{\rm esc} = 0.1$ which causes the reionization to start at $z \sim 20$ and complete at $z \sim 6.5$. The electron scattering optical depth $\tau$ turns out to be 0.07, which is consistent with the recent constraints from Planck \citep{2015arXiv150201589P}. The mass averaged neutral hydrogen fraction, shown in Figure \ref{ionfrc}, attains a value 0.99 and 0.5 around redshift 14 and 8.5 respectively for $S1$.

The value of $f_{\rm esc} = 0.2$ is higher in model $S2$ leading to early reionization completing at $z \sim 8$. The resulting value of $\tau = 0.076$ is higher than that in model $S1$ but is still consistent with the recent Planck constraints \citep{2015arXiv150201589P}. In presence of a mini-quasar like X-ray source at the centre of each galaxy, the evolution of the heated fraction (defined as the volume fraction of regions with $\TK > \TCMB$) is much faster than the evolution of ionization fraction. As shown in Figure \ref{ionfrc}, the universe becomes completely heated below redshift $\sim$ 12 for both the source models $S1$ and $S2$. 

Note that the minimum halo mass for $S1$ and $S2$ source models is $\sim 2 \times 10^9\, \MSUN$, i.e., we have not included any sub-grid sources in these two models. The effect of small mass haloes is studied in model $S3$ where haloes as small as $10^8\, \MSUN$ are included using the sub-grid prescription discussed in \ref{nbody_sim}. The value of $f_{\rm esc}$ is varied at every redshift so as to generate a reionization history (i.e., mass averaged neutral fraction) identical to the model $S2$. In addition, we have included the effects of radiative feedback in model $S3$.

As in paper I, we will consider three different models of heating and $\lya$ coupling and study the light-cone effect. The models are listed in Table \ref{tab1}.
Model A assumes $\TS$ to be much larger than $\TCMB$, this model is identical to models used in \citet{Datta2012b, plante13, datta2014b}. Model B assumes the IGM kinetic temperature to be coupled with the spin temperature through $\lya$ radiation (i.e, the $\lya$ coupling coefficient $x_{\alpha}({\bf x}) \gg 1$), but the heating is calculated self-consistently. Finally in model C, we have calculated the heating and $\lya$ coupling self-consistently and thus it is the most realistic model to be considered while generating $\TB$ maps.  The ionization history in Figure \ref{ionfrc} is same for all the three models A, B and C, while the heating history is same for the models B and C. The effect of peculiar velocities of the gas is also taken into account in all the three models. 

\begin{figure*}
\begin{center}
\includegraphics[scale=0.85]{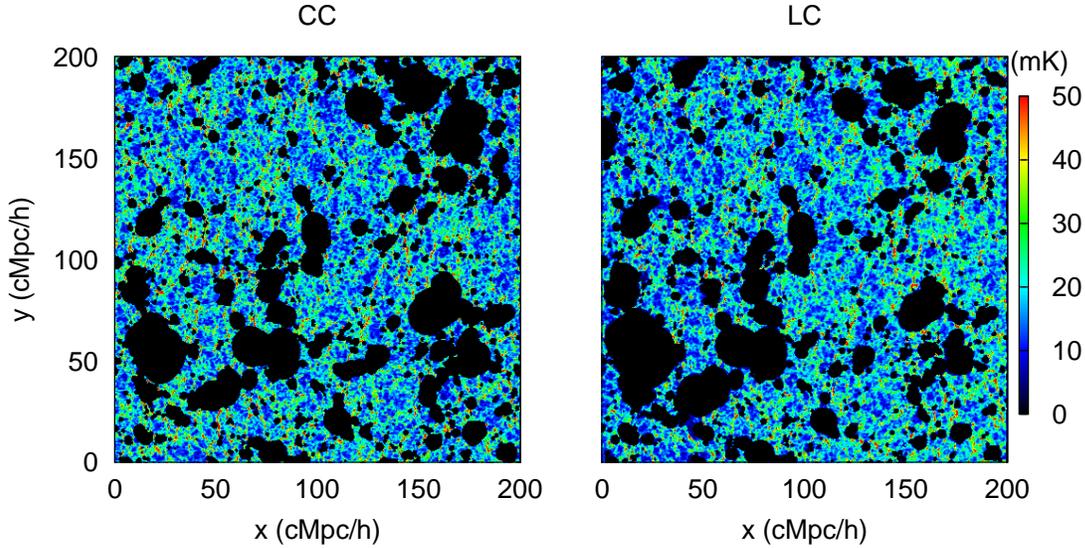}
    \caption{Randomly chosen two-dimensional slice of the $\TB$ distribution with RSD included from the coeval cube (left) and light-cone cube (right) at redshift 9.5 (with mass averaged ionization fraction 0.5) for the model A with the source model $S2$. The comoving size of the box is 200 cMpc $h^{-1}$. The central redshift of  the light-cone (LC) cube is 9.5, whereas the redshift span from left to right along the x-axis (LOS direction) is 8.86 to 10.13. Note that this model makes the assumption $\TS >> \TCMB$ and is driven by haloes identified using spherical overdensity halo-finder in the simulation box.}
   \label{tb_slice_a}
\end{center}
\end{figure*}

\begin{figure*}
\begin{center}
\includegraphics[scale=0.85]{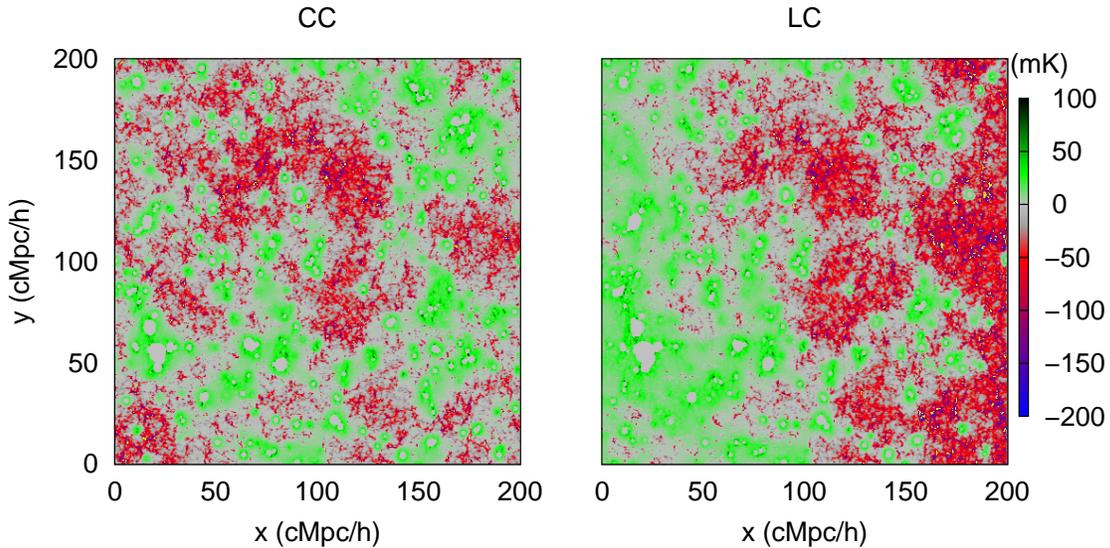}
    \caption{Similar to Figure \ref{tb_slice_a}, but for the model C and the coeval cube corresponding to redshift 13 when the universe was 50 \% heated by volume for the source model $S2$. The left and right hand edge of the `LC' box correspond to redshifts 12 and 14 respectively.}
   \label{tb_slice_c}
\end{center}
\end{figure*}

\begin{table}
\centering
\small
\tabcolsep 3pt
\renewcommand\arraystretch{1.2}
   \begin{tabular}{c c c c c c c}
\hline
\hline
    Source   &$f_{\rm esc}$  & $\tau$ & $z_{\rm end}$	& $M^{\rm min}_{\rm halo}$ & sub-grid & radiative\\
      model                                                      &&&&($\MSUN$)& halo & feedback\\
\hline
\hline
    $S1$      & 0.1        & 0.07    	    &	6.5		& $2\times 10^9$ & no	& no	\\
    $S2$      & 0.2        & 0.076	    & 7.8   &	$2\times 10^9$  & no  & no \\
		$S3$    	& varying      & 0.076   	& 7.8		&	$10^8$	& yes & yes	\\
\hline
\end{tabular}
\caption[]{Properties of the source models considered in this paper. The UV escape fraction $f_{\rm esc}$ for the model $S3$ is varied to get similar ionization history like $S2$.}
\label{tab2}
\end{table}

\begin{table}
\centering
\small
\tabcolsep 3pt
\renewcommand\arraystretch{1.2}
   \begin{tabular}{c c c}
\hline
\hline
    Model   &$\lya$ coupling  & Heating  \\
\hline
\hline
    A      &Coupled        & Heated    \\
    B			&Coupled      & Self-consistent   \\
		C    &Self-consistent      & Self-consistent   \\
\hline
\end{tabular}
\caption[]{Properties of models considered to investigate the effects of inhomogeneous heating and $\lya$ coupling on the signal. The terms `coupled' and `heated' represent the scenarios where $x_{\alpha} \gg 1$ and  $\TK$ $\gg$ $\TCMB$ respectively.}
\label{tab1}
\end{table}


\subsection{Visualizing the light-cone effect}
\label{res_vis}

\begin{figure*}
\begin{center}
\includegraphics[scale=0.65]{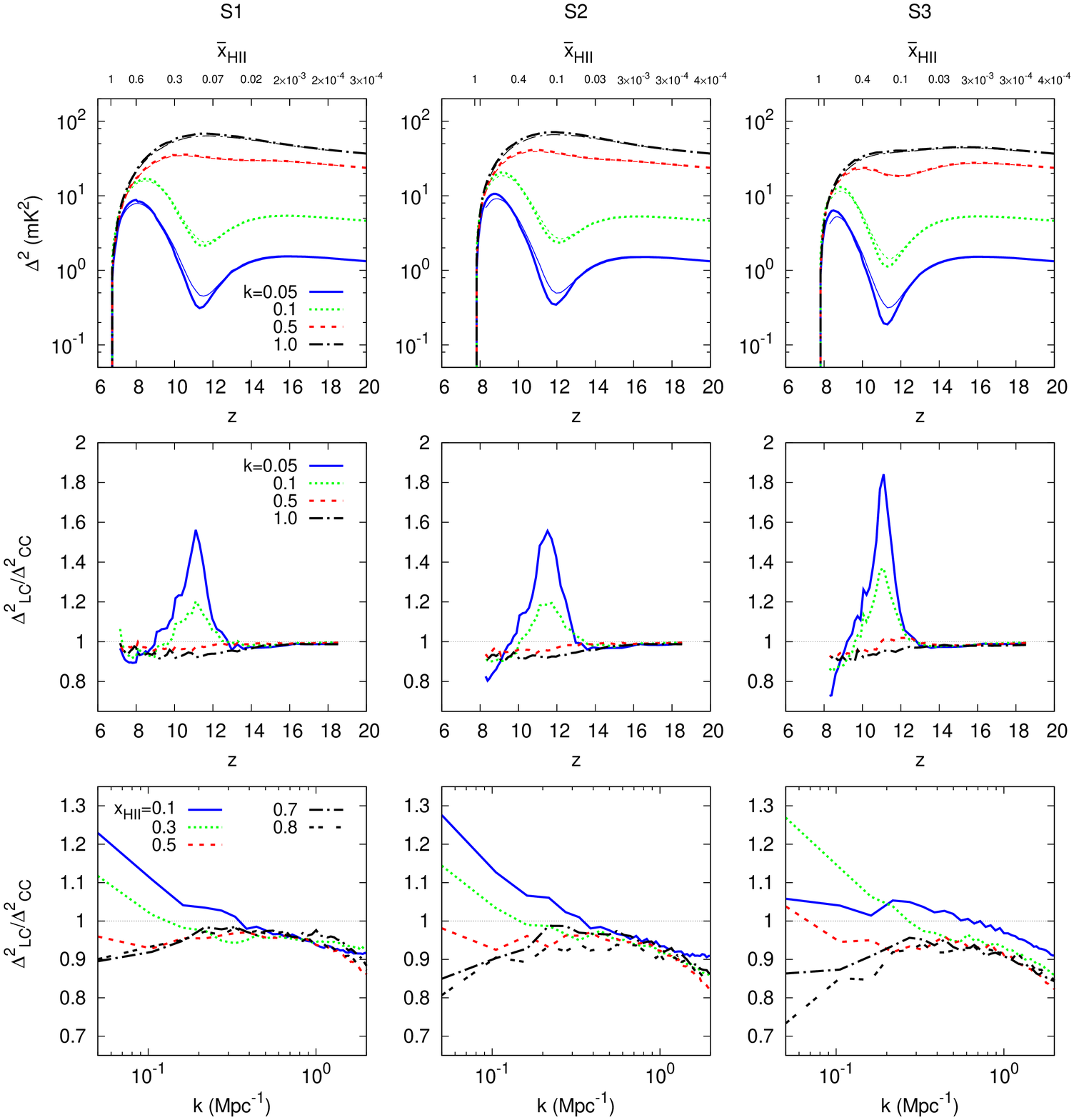}
    \caption{Model A: The light-cone effect on the power spectrum of 21-cm brightness temperature fluctuation. The upper most panels show the spherically  averaged power spectrum of $\TB$ with (thin curves) and without (thick curves) light-cone effect as a  function of redshift for four different scales $k/$Mpc$^{-1} = 0.05$ (solid), 0.1 (dotted), 0.5 (dashed) and 1.0 (dot-dashed ). The middle row panels show the ratio of the power spectra with and without light-cone effect as a function of redshift. The bottom most panels show the ratio of the power spectra with and without light-cone effect as a function of scales for different ionization fractions (0.1,0.3, 0.5, 0.7 and 0.8). This model makes the assumption that $\TS >> \TCMB$. Three columns represent three different source models $S1, S2$ and $S3$ respectively. Redshift space distortion is included in all the panels.}
   \label{tb_modela}
\end{center}
\end{figure*}

The light-cone effect can be easily visualized in the maps of $\TB$ distribution for different models. Figure \ref{tb_slice_a} shows the $\TB$ map computed on a randomly chosen slice from the simulation box for the model A with the source model $S2$. The left hand panel shows the distribution of $\TB$ from the coeval cube  (with redshift space distortion included, but without the light-cone effect) at redshift $9.5$ with mass averaged ionization fraction $0.5$. The right hand panel shows the same slice with the light-cone effect incorporated. Along the line of sight (horizontal axis), the central redshift of  the light-cone (LC) cube is $9.5$, whereas the light-cone cube spans from redshift $8.86$ (left edge of the box) to $10.13$ (right edge of the box). 
The ionization bubbles (black regions) are larger at the left-hand side of the middle of light-cone cube as compared to the coeval cube. This is because of the fact that the left-hand side regions correspond to lower  redshifts than the middle region of the box and thus correspond to later stages of reionization where the bubbles are of systematically larger. It is opposite in the case of right-hand side regions. 

Figure \ref{tb_slice_c} shows similar slices at redshift 13 for the model C. In this case not only the ionized regions appear larger in the front side of the light-cone cube compared to the coeval cube, but the heated regions (regions with $\TK > \TCMB$) too show similar trend. One should notice that the left half shows the signal predominantly in emission whereas the right half shows the same mostly in absorption, although the corresponding coeval slice shows that emission and absorption regions are homogeneously distributed over the entire slice. This significant imbalance of the emission regions between the left and right-hand regions of the light-cone cube is expected to introduce effects on the observable signal at initial stages of reionization.


\subsection{Comparison with previous studies}
\label{res_com}

It is possible to validate our formalism by comparing the model A where we have assumed $\TS >> \TCMB$ with the existing works in the context of the light-cone effect on the 21-cm signal. Figure \ref{tb_modela} shows the light-cone effect on the spherically averaged power spectrum of the 21-cm $\TB$ fluctuations (with redshift space distortion included) for the model A.\footnote{Note that we  exclude modes with ${\bf k_{\bot}}=0$ while calculating the power spectrum, as these modes are not accessible to the interferometers.} The upper panels show the evolution of the dimensionless power spectrum for the three source models as a function of redshift for four different scales $k$ = 0.05, 0.1, 0.5 and 1 Mpc$^{-1}$ which are represented by the solid, dotted, dashed and dot-dashed curves, respectively. The signal is always in emission for this model and thus the power spectra have relatively smaller amplitudes ($\sim 10\, \rm mK^2$ ) at large scales. The large scale power spectrum for each of the three source models shows one trough and one peak when plotted as a function of redshift around the period when universe was $\sim 10\%$ and $\sim 65\%$ ionized by mass. Initially the highly dense regions get ionized and thus \HI ~fluctuations at large scales are suppressed. This results in the trough like feature in the evolution of the power spectrum \citep[e.g,][]{Majumdar13,datta2014b}. The power spectrum then increases with the growth of bubble size and finally decreases with the decrease in ionization fraction, which generates the prominent peak around ionization fraction 0.65 for this model.

 The thick (thin) curves in the upper panels of the figure represent the power spectra without (with) the light-cone effect included. It is clear that the light-cone effect is most substantial around the trough and the peak regions in the evolution of the power spectra. Because of this effect, the power spectra can be enhanced by a factor as large as $\sim 50$\% (when $\XHII \sim$ 0.15) and suppressed by $\sim 20$\% (when $\XHII \sim$ 0.8) for the model $S2$ at large scales $k = 0.05\, \rm Mpc^{-1}$, as shown in the middle panels of Figure \ref{tb_modela}. The effect is minimum around the period when universe is $\sim 50 \%$ ionized. This is because of the fact that around $50\%$ ionization, any linear evolution of the power spectrum as a function of $z$ is smoothed out, as was pointed out by \citet{Datta2012b}. On the other hand, when $\XHII \sim$ 0.1 and 0.8 the evolution of the power spectrum is highly non-linear which makes the effects strong \citep{Datta2012b,datta2014b}. We also see that the effects are less for a smoother ionization model like $S1$ compared to $S2$. The reason is that in case of $S1$, the evolution of the ionized bubbles are less rapid compared to model $S2$. The evolution of power spectrum is more non-linear at the trough and the peak regions in presence of small mass haloes as shown in the upper panel for the model $S3$. Thus the light-cone effect in model $S3$ is larger compared to the other two models. The bottom most panels of the figure show the scale dependence of the ratio of the power spectrum with and without light-cone effect at different stages of the reionization history. The evolution of the power spectrum at small scales is much more linear as shown in the upper panels of the figure and thus the light-cone effect is smaller at small scales compared to larger scales. Our results for the model A are consistent with a similar model presented in \citet{datta2014b}.


\subsection{Effect of inhomogeneous heating and $\lya$ coupling on the light-cone effect}
\label{res_lc}

\begin{figure*}
\begin{center}
\includegraphics[scale=0.65]{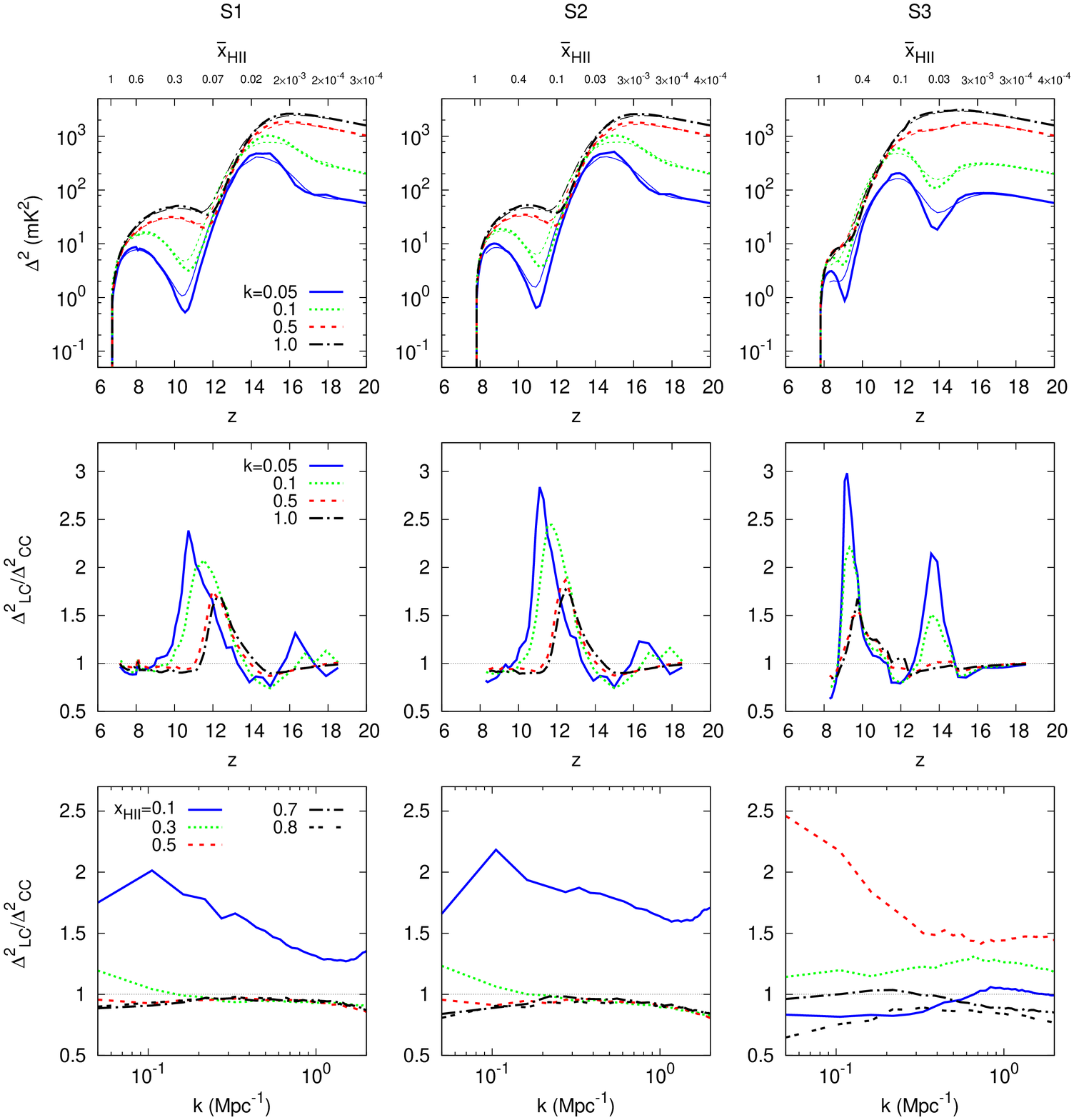}
    \caption{ Same as Figure \ref{tb_modela}, but for the model B where the IGM is assumed to be $\lya$ coupled while the heating is calculated self-consistently.}
   \label{tb_modelb}
\end{center}
\end{figure*}

We now discuss the light-cone effect in presence of non-uniform heating and $\lya$ coupling. Figure \ref{tb_modelb} shows the same quantities plotted in Figure \ref{tb_modela} but  for the model B (where the fluctuations in heating are calculated self-consistently, but the $\lya$ coupling process is assumed to be very efficient all throughout). The top panels show the evolution of power spectra for different length scales. One can compare the plots with those in Figure \ref{tb_modela} and immediately conclude that the evolution is much more rapid in the model B. For example, the amplitude of the power spectrum is $\sim 50$ mK$^2$ at $z \sim 20$ which increases up to $\sim 500$ mK$^2$ at $z \sim 15$ and then decreases rapidly to $\sim 0.5$ mK$^2$ at $z \sim 11$ for $k = 0.05$ Mpc$^{-1}$ (in the $S2$ model). The corresponding evolution is much less rapid for the model A where the amplitude is almost constant at $\sim 1$  mK$^2$ from $z = 20$ till $z \sim 14$. The light-cone effect is thus expected to be more significant for the model B.
 
 The evolution of the power spectrum is faster in model B, due to the fact that the heated bubbles are larger and grow much faster than the ionized bubbles. This causes the evolution of the heated fraction to be very rapid compared to evolution of the ionization fraction (see Figure \ref{ionfrc}). As a large fraction of the gas in the IGM has temperature $\TK$ less than $\TCMB$ during cosmic dawn and initial phase of reionization (as gas temperature falls as $(1+z)^2$ due to adiabatic expansion of the universe after the decoupling around redshift $z_{\rm dec}\sim150$ and $\TCMB$ falls as $(1+z)$) and the model assumes $\TS=\TK$, the model shows very large absorption signal from the cosmic dawn until the IGM is sufficiently heated to follow the model A from $z\sim10$ for $S2$. The inhomogeneous X-ray heating from the mini-quasars results in increasing the $\TB$ fluctuations, which produces a distinct peak in the power spectrum at large scales around $z\sim 15$ when plotted against $z$. The second prominent peak occurs at $z \sim 9$ because of ionization fluctuations analogous to the model A. The amplitude of the power spectrum at the heating peak ($\sim 10^{3}\, \rm mK^{2}$) is significantly higher compared to the ionization peak ($\sim 10\, \rm mK^{2}$)  and hence this redshift range might be of interest to interferometers like the SKA. The reason for this is that at the ionization peak around $z \sim 9$, the power spectrum is dominated by ionization fluctuations and the signal is in emission because the heating is already substantial. On the other hand, at the heating peak around $z \sim 15$, the $\TB$ field consists of both emission and absorption regions, and the power spectrum is dominated by $\TS$ fluctuations (Paper I).

\begin{figure*}
\begin{center}
\includegraphics[scale=0.65]{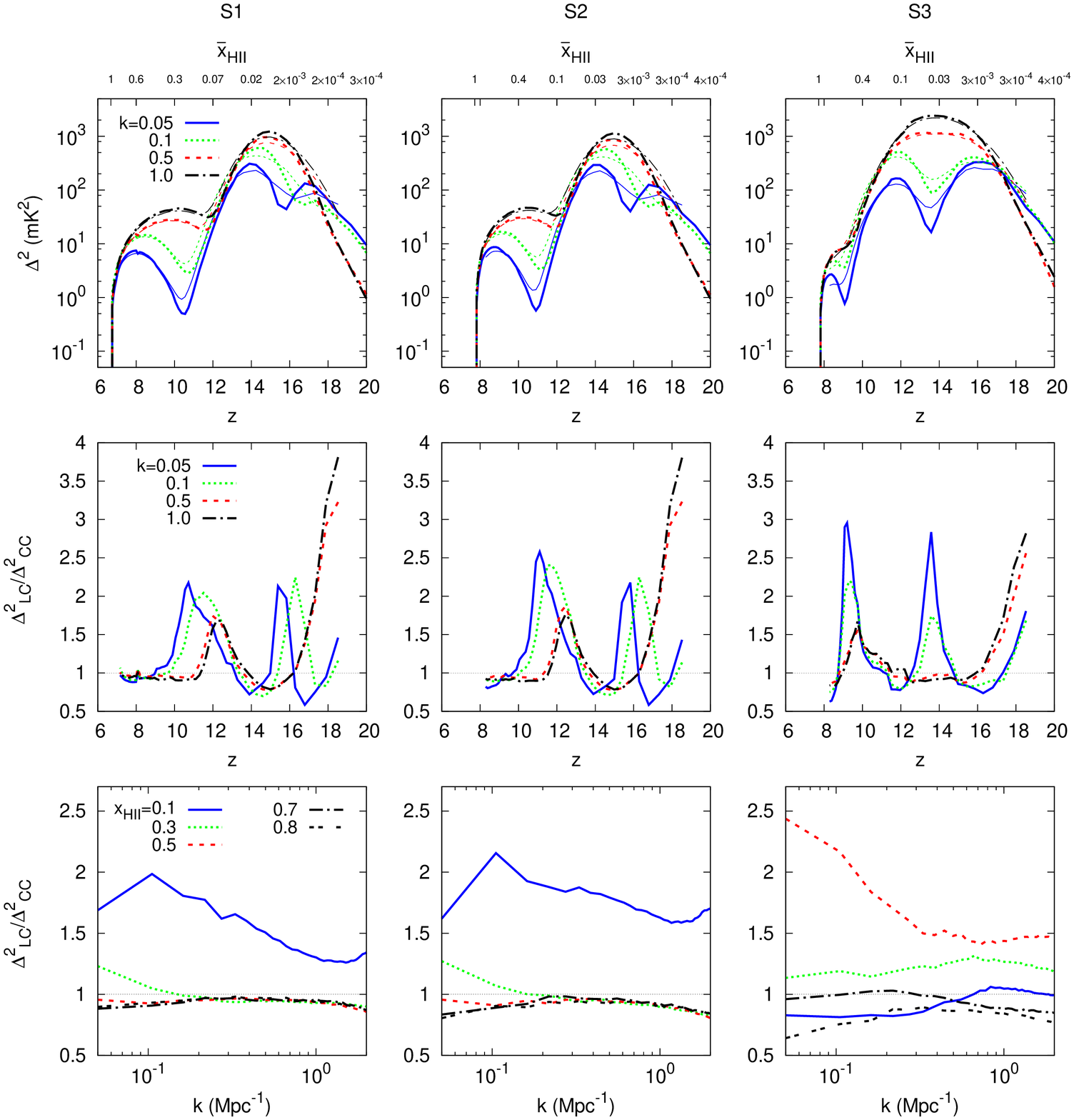}
    \caption{ Same as Figure \ref{tb_modelb}, but for the model C where both $\lya$ coupling and heating are calculated self-consistently.}
   \label{tb_modelc}
\end{center}
\end{figure*}

\begin{figure*}
\begin{center}
\includegraphics[scale=0.65]{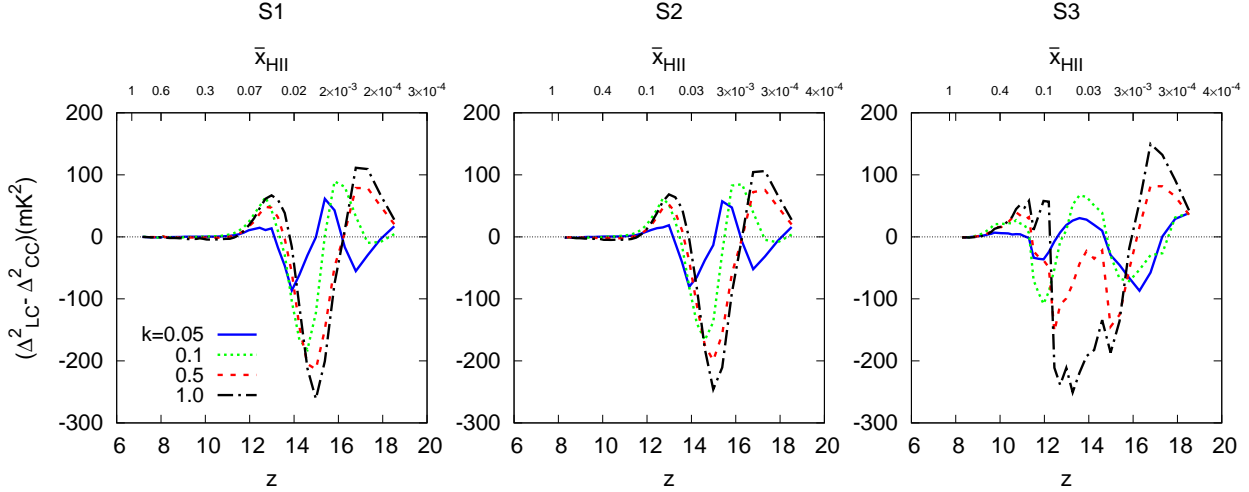}
    \caption{The redshift evolution of the difference between the power spectrum with and without  light-cone effect at scales $k/$Mpc$^{-1} = 0.05$ (solid), 0.1 (dotted), 0.5 (dashed) and 1.0 (dot-dashed) for the model C. Left to right panels represent source models $S1, S2$ and $S3$ respectively.}
   \label{lc_diff}
\end{center}
\end{figure*}

As can be seen from Fig \ref{tb_modelb}, the light-cone effect is much more prominent in the model B compared to the model A. When the light-cone effect is included in the analysis, we find maximum suppression of power spectrum around the two peak regions and increase around the trough region between the two peaks. Inhomogeneous heating makes the evolution of the power spectrum much more non-linear  and results in a stronger light-cone effect on the power spectrum than that in the model A. The figure also shows that the light-cone effect can increase the power spectrum by a factor as large as 1.23 and 2.84 times around redshifts 16 and 11 respectively at large scales ($k \sim 0.05\, \rm Mpc^{-1}$) for the model $S2$. The suppression of the power spectrum around the peaks around redshifts 8 and 15 can be $\sim 20-25 \%$ at large scales. The redshift evolution of the power spectrum at small scales is almost linear during the cosmic dawn and initial phase of reionization as shown in the upper most panels of the figure and thus the light-cone effect is minimum at small scales during these epochs. Around redshift 12 when the IGM becomes highly heated, the evolution in the small scale $\TB$ fluctuations becomes extremely fast which results in the increment of the power spectrum by a factor of 1.5 due to the light-cone effect at small scales ($k \sim 1\, \rm Mpc^{-1}$). The light-cone effect is less in the model $S1$, as the evolution of the ionized as well as the heated regions is less rapid compared to that in the model $S2$.  In presence of small mass haloes in the model $S3$, the heating is delayed compared to the model $S2$ as we have varied $f_{\rm esc}$ and kept $f_X$ fixed. In this case, the fluctuations in $\TB$ at large scales is smaller compared to that in the model $S2$ and thus the amplitude of the heating peak is smaller. As the heated regions around the sources are very small in $S3$ and eventually increase, the power spectrum shows a dip in its evolution with the redshift as shown in the top-right panel of the figure. The light-cone effect is able to increase the power spectrum $\sim 2$ times around the dip (at $z \sim 14$).  

The amplitude and overall nature of the heating peak around redshift 15 depend on the nature of X-ray sources. As for example, the peak may not present if heating is driven by X-ray sources like high mass X-ray binaries (HMXBs) as they do not have large number of soft X-ray photons and thus the heating will be almost homogeneous \citep[e.g,][]{Fialkov14}. In such cases, $\TB$ fluctuations will be dominated by the neutral hydrogen fluctuations if we consider the $\lya$ coupling to be very high from the beginning. Thus the light-cone effect in such scenarios will be similar to the model A. Whereas, for X-rays, from sources like supernova, hot-interstellar gas, follow similar profile like mini-quasar and thus expected to have similar light-cone effect. 

\begin{figure*}
\begin{center}
\includegraphics[scale=0.65]{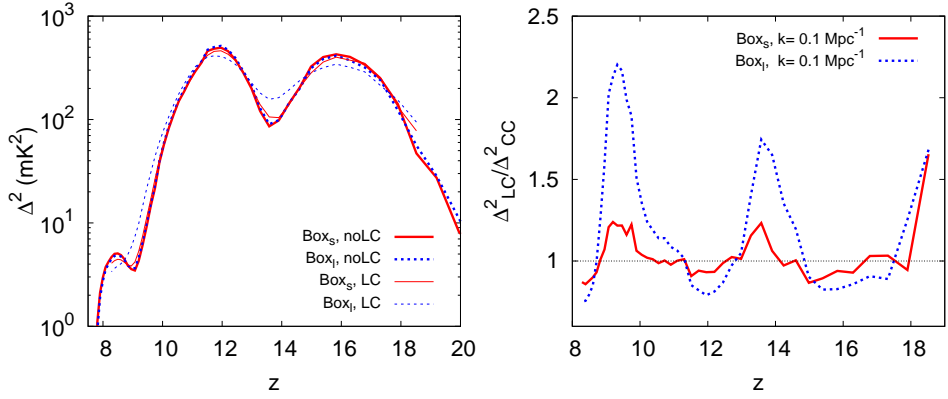}
    \caption{The dependence of the light-cone effect on the simulation box size. The left hand panel shows the evolution of the power spectrum at a scale $k = 0.1$ Mpc$^{-1}$ for  the two different boxes $\rm Box_l$ ($200\, h^{-1}\, \rm cMpc$, blue-dotted curves) and $\rm Box_s$ ($100\, h^{-1}\, \rm cMpc$, red-solid curves) respectively with (thin) and without (thick) the light-cone effect. The right hand panel shows the redshift evolution of the ratio of the power spectra with and without the light-cone effect. The plots are for $S3$ model and redshift space distortion effect is included while calculating the power spectrum.}
   \label{small_box}
\end{center}
\end{figure*}

\begin{figure*}
\begin{center}
\includegraphics[scale=0.65]{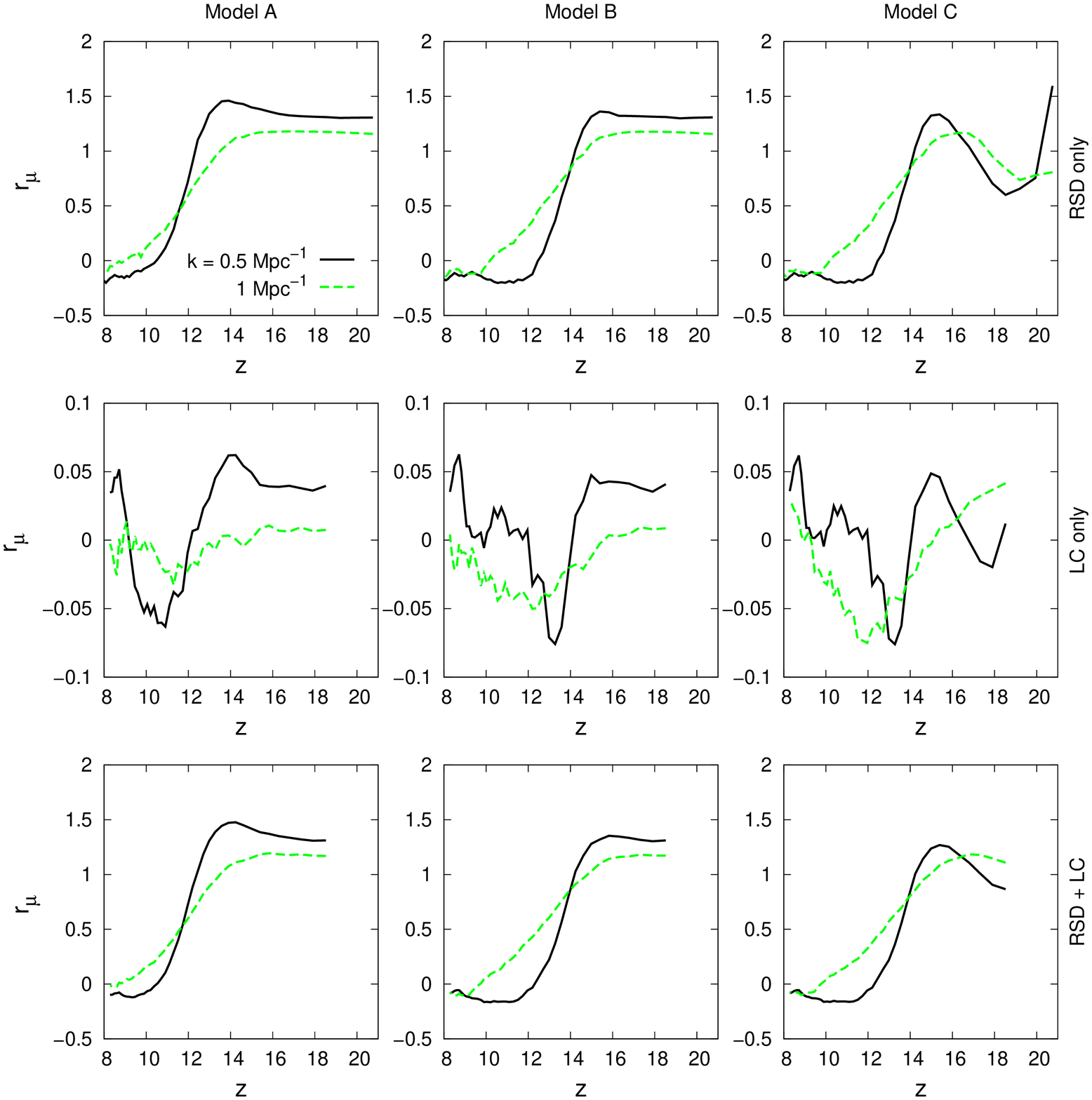}
    \caption{Evolution of the `anisotropy ratio' $r_{\mu}$ as a function of redshift for the source model $S3$ at two different scales $k=0.5, 1\, \rm Mpc^{-1}$. The three columns from left to right represent models A, B and C respectively. The top (middle) row represents the case where only the RSD (light-cone) effect is included, while the bottom most row represents the case where both the effects are included.}
   \label{ani_r_ratio}
\end{center}
\end{figure*}

The difference between models B and C is that the $\lya$ coupling is calculated self-consistently in model C whereas in model B it is assumed that the IGM is $\lya$ coupled. As the condition $x_{\alpha}({\bf x}) \gg 1$ is not valid for major part of the IGM during the cosmic dawn, the amplitude of $\TB ({\bf x})$ in the model C will be much less than the predicted $\TB$ in the model B. As a result the amplitude of the power spectrum is much lower than in the model B initially.    The inhomogeneous $\lya$ coupling during the cosmic dawn in the model C increases the fluctuations in $\TB$ and results in a distinct peak in the evolution plot of the large scale power spectrum as a function of redshift as shown in  the upper panels of Figure \ref{tb_modelc}. We note that a very small amount of $\lya$ photons is sufficient to couple $\TS$ with $\TK$ of the IGM, thus the model C follows the model B very soon the first sources formed (around redshift 13).	

In presence of inhomogeneous $\lya$ coupling the evolution of the power spectrum is more dramatic at both large and small scales at cosmic dawn. The non-linear rise of the power spectrum, from $\sim 1\, \rm mK^2$ to $\sim 100\, \rm mK^2$ within redshift interval 20 to 16, results in a stronger light-cone effect at the very beginning of the reionization epoch. This increases the power spectrum by a factor of $\sim 1.5$ around redshift 18.5 and suppresses by a factor of 0.6 around redshift 16.5 (which corresponds the $\lya$ peak for the source model $S2$). The light-cone effect further enhances the power spectrum by a factor of $\sim 2$ around redshift 15.5 (trough region between the $\lya$ and heating peaks), followed by a suppression by a factor of 0.7 around redshift 14.  For redshift $z < 14$, the light-cone effect is similar to the model B, only difference being that in presence of inhomogeneous $\lya$ coupling the effect is little weaker than that predicted by the model B.  The power spectrum evolves rapidly at small scales at the cosmic dawn, which results in a large light-cone effect at small scales ($\sim 3-4$ times enhanced). As expected, the light-cone effect is found to be smaller in model $S1$ compared to model $S2$. The enhancement in the power spectrum due to the light-cone effect is $\sim 3$ around redshift 14 which corresponds to the trough region between the $\lya$ coupling and heating regions for the model $S3$.

Importantly, we also notice that the peaks and dips found in the evolution of the power spectra for the models B and C are smoothed out to some extent due to the light-cone effect (upper panels of Figs \ref{tb_modelb} and \ref{tb_modelc}). The effect is more prominent at large scales $k \lesssim 0.1$ Mpc$^{-1}$. The light-cone effect lowers the peak height and raises the dip by some amount.  This is particularly important for $S3$ in model C, which is probably the most realistic reionization model we consider here. It has been suggested \citep[e.g.,][]{Mesinger2014} that the peaks can be used to extract source properties, put constrain on the X-ray and $\lya$ background during the cosmic dawn. We argue that the light-cone effect should be considered while extracting those parameters from the peak and dip measurements. 

Finally, in Figure \ref{lc_diff} we show the redshift evolution of the difference in the power spectra  with and without the light-cone effect for the model C  for the three source models at different scales. As expected, the difference is maximum at the peak and the trough regions found in the evolution of the power spectra. At large scales ($k \sim 0.05\, \rm Mpc^{-1}$), the difference can be as large as $\sim -100 $ to $100~ \rm mK^2$ . The difference is $\sim -250 $ to $100~ \rm mK^2$ at intermediate scales like  $k \sim 1\, \rm Mpc^{-1}$. In principle, such a strong effect should easily be detected by future experiments like the SKA.


\subsection{Effect of simulation box size}
\label{sim_box}

It has been found earlier by \citet{datta2014b} that the light-cone effect is larger when larger simulation box is used. In order to quantify the effect of box size on our results, we estimated the signal using a smaller box of length $100\, h^{-1}$ cMpc (in addition to our default box of $200\, h^{-1}$ cMpc). The results are shown in Fig. \ref{small_box}. In the left panel, we plot the evolution of $\Delta^2$ at a scale $k = 0.1$ Mpc$^{-1}$ for the two boxes with and without the light-cone effect.  As expected, the power spectra without the light-cone effect  (thick lines) for the two boxes agree with each other. However, we can see that the light-cone effect is considerably less prominent for the smaller box in agreement with \citet{datta2014b}. In particular, the smoothing of the three peak nature in the evolution plot of the power spectrum is more prominent for the larger box. The effect can be much more stronger in simulation box with size $\sim 600 $ cMpc \citep[e.g.,][]{datta2014b, Mesinger2014}, which may completely smooth out the three peak nature of the evolution plot of the power spectrum. This will constrain us to choose smaller frequency bands during 21-cm experiments to avoid strong light-cone effects and restore the peakiness feature of the power spectrum which is very useful for parameter estimation etc. The same conclusions can be drawn from the right hand panel too where we have shown the ratio of the power spectra with and without the light-cone effect for $k = 0.1$ Mpc$^{-1}$. Clearly, the ratio deviates from unity quite prominently for the larger box compared to the smaller one.


\subsection{Anisotropy from the light-cone effect}
\label{ani_lc}

Since the light-cone effect modifies the 21-cm signal along the line of sight direction similar to the RSD, it is expected that it may cause anisotropies in the signal. This was investigated in detail by \citet{datta2014b} for a model similar to our model A, and they concluded that the light-cone effect does not induce any significant anisotropies in the signal. We confirm their findings for the model A. In addition, we find that the light-cone effect does not cause any prominent anisotropies for the models B and C too for relevant scales of interest. In fact, for large scales $k \lesssim 0.1$ Mpc$^{-1}$, our simulation box does not contain sufficient number of modes leading to large sample variance. Hence it is difficult to draw any significant conclusion on anisotropies at large scales.

In order to explain the effect in a somewhat simpler manner, we have calculated the `anisotropy ratio' defined as  \citep[e.g.,][]{2015PhRvL.114j1303F}:
\begin{equation}
r_{\mu}(k,z)=\frac{\left<\DEL2({\bf k},z)_{|\mu_{k}|>0.5}\right>}{\left<\DEL2({\bf k},z)_{|\mu_{k}|<0.5}\right>}-1,
\label{ani_r_define}
\end{equation}
where the averages are over angles. If the signal is isotropic then $r_{\mu}(k,z)$ will be identically zero, otherwise it can be positive or negative. The redshift evolution of the anisotropy ratio  for scales $k = 0.5, 1$ Mpc$^{-1}$ is plotted in Figure \ref{ani_r_ratio}. The source model chosen is $S3$. We find from the top panels that the RSD can cause significant anisotropies ($r_{\mu} \sim 1.5$ for $k \sim 0.5$ Mpc$^{-1}$) for all the three models\footnote{For the models B and C, the level of anisotropies arising from the RSD depends on the source model too. The anisotropies are caused by the correlation of the $\TB$ fluctuations with the dark matter density field, which is enhanced when small sources are included. For $S1$ and $S2$ which do not contain haloes smaller than $\sim 10^9 \MSUN$, we find that $|r_{\mu}| \lesssim 1$ from RSD effects (in agreement with results of Paper I).}. The redshift where the $r_{\mu}$ is maximum corresponds to prominent dip in the evolution of the power spectrum.

In contrast, the light-cone effect does not cause any significant anisotropy on the signal as can be seen from the middle panels for the scales considered. The anisotropy, when both the effects are included, is thus dominated by the contribution from the RSD. We should mention here that it is possible that the anisotropy arising from light-cone effect behave differently for larger scales $k \lesssim 0.1$ Mpc$^{-1}$. In order to have sufficient number of modes at such large scales one requires boxes of much larger size which, unfortunately, are beyond the scope of this paper.

\section{Summary and discussion}
\label{conc}

The main focus of the paper is to investigate the impact of the light-cone effect on the \HI ~21-cm signal from the cosmic dawn. The 21-cm brightness temperature maps are generated by post-processing a dark matter $N$-body simulation with an 1-D radiative transfer code. In addition to the usual stellar-like sources, we have accounted for X-ray emitting mini-quasar like sources. The fluctuations in the spin temperature due to inhomogeneous $\lya$ coupling and heating of the IGM not only boost the 21-cm power spectrum but introduce several  peaks and dips in it when plotted as a function of redshift. The boosted power spectrum, together with additional peaks and dips along the redshift axis, makes it an ideal case for studying the light-cone effect. Here, we calculate the spin temperature fluctuations self consistently for various possible source models and reionization histories. We then study, for the first time, the light-cone effect on the 21-cm power spectrum arising from the cosmic dawn when the signal is dominated by spin temperature fluctuations (unlike other previous studies which concentrated mainly on the ionization fluctuations).

The main findings of our work are summarised below:

\begin{enumerate}

\item We find that our results are consistent with previous studies \citep{datta2014b} for the model which assumes that the \HI ~spin temperature to be fully coupled with the IGM temperature and very high compared to the CMB temperature throughout the reionization history (model A). The light-cone effect is most significant when the universe is $\sim 15\%$ and $80\%$ ionized as the spherical averaged power spectrum is increased/suppressed by a factor of $\sim 1.5$ and $\sim 0.8$ respectively. The effect is found to be minimum while the universe is $\sim 50\%$ ionized.

\item The light-cone effect is much more dramatic in the model B where the inhomogeneous heating is taken into account. We find that the maximum suppression of the large-scale power spectrum due to the light-cone effect occurs around the two peaks and enhancement occurs around the dip between the two peaks when plotted against $z$. We notice that the light-cone effect can increase the power spectrum by a factor of $\sim 1.25$ and $\sim 3$ times around redshift $16$ and $11$ respectively at large scales ($k \sim 0.05 \, {\rm Mpc}^{-1}$). The suppression of the power spectrum around the peaks around redshifts $8$ and $15$ can be $\sim 20$ to $25\%$ at large scales. We find the enhancement/suppression  to be higher for models where contributions of small sources have been included ($S3$ source model). Unlike the previous studies, we find a significant light-cone effect at small scales ($k \sim 1 \,{\rm Mpc}^{-1}$) as well at high redshifts.

In addition, when we calculate the $\lya$ coupling self-consistently (model C), we find that the power spectrum increases very rapidly at the very beginning of cosmic dawn and thus the light-cone effect can enhance the 21-cm power spectrum by a factor $\sim 1.5$ at large scales ($k \sim 0.05\, \rm Mpc^{-1}$). At these scales, the three peaks of the power spectrum are suppressed by factors of $0.6$, $0.75$ and $0.8$ around redshifts $16.5, 14$ and $8.5$ respectively. The light-cone effect can enhance the power spectrum in the dips by factors of $2$ and $2.5$ around redshift $15.5$ and $11$ respectively. In general, the light-cone effect at small scales is found to be smaller than the effect at larger scales, except during the cosmic dawn (may increase the power spectrum by  factor of few at scales like $k \sim 1\, \rm Mpc^{-1}$).

\item Our results can be understood from the fact that the light-cone effect is strong in a situation where the power spectrum evolve non-linearly with redshift (or comoving distance) as any linear evolution gets cancelled out to a large extent \citep{datta2012a}.  Non-linear evolution of the 21-cm power spectrum is maximum when it takes a turn. For example, for the A model it happens twice, first time when $x_{\rm HI} \sim 0.8$ and second time when $x_{\rm HI} \sim 0.3-0.4$.  These are the places where the light-cone effect is very strong. For the B type model, one more peak in the power spectrum gets added.  Because of this non-linear evolution around the heating peak, we see that the light-cone effect becomes very strong (power spectrum is suppressed substantially) around this peak.  Similarly $\lya$ coupling inhomogeneities give rise to two more episodes of non-linear evolution of the power spectrum (one dip and one peak) in the very beginning of cosmic dawn where the light-cone effect is very strong.

\item The large scale ($k \sim 0.05\, \rm Mpc^{-1}$) power spectrum  with light-cone effect included can differ by $\sim -100$ to $100$ mK$^2$ from the case where light-cone effect is not included. The difference increases at small scales ($k \gtrsim 0.5\, \rm Mpc^{-1}$) to the range $\sim -250$ to $100$ mK$^2$. In principle, such a strong effect should easily be detected by future experiments like the SKA. We also notice that the peaks and dips in the power spectrum are somewhat smoothed out due to the light-cone effect. It has been suggested \citep[e.g.,][]{Mesinger2014} that the peaks can be used to extract source properties, X-ray and $\lya$ background etc. We argue that the light-cone effect should be considered while extracting those parameters from the peak and dip measurements.

\item The light-cone effect can, in principle, introduce anisotropies in the power spectrum, similar to the RSD effects. It is somewhat difficult to predict the large scale $k \lesssim 0.1$ Mpc$^{-1}$ anisotropies because of the lack of number of modes in the simulation box. However, for intermediate scales $k \sim 0.5$  Mpc$^{-1}$, we do not find any significant anisotropy arising from the light-cone effect.

\end{enumerate}

The light-cone effect, during the cosmic dawn ($x_{\rm HII} \lesssim 0.2)$ is highly  dependent on nature of the X-ray sources present during that time in the universe. For example, the heating of the IGM will be  much more homogeneous if it is mainly driven by the HMXBs \citep{Fialkov14}. In such a scenario the evolution of the signal will be solely dominated by the \HI ~neutral fraction and thus the light-cone effect during the cosmic dawn will not be as strong as we see here. Although, we expect similar light-cone effect if supernova or hot inter-stellar gas dominate the X-ray budget since they are expected to have similar kind of SED like mini-quasars which emit large number of soft X-ray photons. 

Finally, further investigation is required in order to understand the light-cone effect for different source models and reionization scenarios. The main result from this study is that the light-cone effect can change the amplitude of the 21-cm power spectrum as well as the shape of the evolution of the power spectrum at large scale quite substantially. Therefore, it is important  that the effect is  incorporated  while modelling the 21-cm power spectrum, particularly  during the cosmic dawn and early stages of reionization.

\section*{Acknowledgement}

The simulations used in this work were performed on the IBM cluster hosted by the National Centre for Radio Astrophysics, Pune, India. KKD would like to thank DST for support through the project SR/FTP/PS-119/2012 and the University Grant Commission (UGC), India for support through UGC-faculty recharge scheme (UGC-FRP) vide ref. no. F.4-5(137-FRP)/2014(BSR).

\bibliography{lc_effect}
\bibliographystyle{mn2e}

\label{lastpage}
\end{document}